\documentclass[runningheads]{llncs}
\usepackage{cite}
\usepackage[T1]{fontenc}
\usepackage{amsmath,amssymb,amsfonts}
\usepackage{algorithmic}
\usepackage{graphicx,color}
\usepackage{booktabs}
\usepackage{multirow}

\usepackage{textcomp}
\usepackage{graphicx} 

\usepackage{floatrow} 
 \usepackage{xcolor}

\usepackage{multirow}
\usepackage{hyperref}
\usepackage{chemformula}
\usepackage{todonotes}

\begin{document}

\title{Analysis of the carbon footprint of HPC}

%
\author{Abdessalam Benhari\inst{1,2} \and Yves Denneulin\inst{1} \and Frédéric Desprez\inst{3}\and Fanny Dufossé\inst{1} \and Denis Trystram\inst{1}}
\institute{Univ. Grenoble Alpes, Grenoble INP, Inria, CNRS, LIG, France \and
Bull SAS (Eviden, Atos group), France
\and Inria, Grenoble, France}
%
\maketitle
%


\begin{abstract}


The demand in computing power has never stopped growing over the years. Today, the
performance of the most powerful systems exceeds the exascale. Unfortunately, this growth also comes with ever-increasing energy costs, leading to a high carbon footprint. This paper investigates the evolution of high performance systems in terms of carbon emissions. 
A lot of studies focus on Top500 (and Green500) as the \textit{tip of an iceberg} to identify trends in the domain in terms of computing performance. 
We propose here to go further in considering the whole span life of several large scale systems and to link the evolution with trajectory toward 2030.
More precisely, we introduce the energy mix in the analysis of Top500 systems and we derive a predictive model for estimating the weight of HPC for the next 5 years.

\end{abstract}

\section{Introduction}
\label{sec:intro}


With the climate crisis and global warming, humanity faces an unprecedented challenge~\cite{IPCC2022}. 
It is well-established that the ICT sector (Information and Communication Technologies) plays a significant role in this crisis through the huge energy consumption of computing, storage, and interconnection devices. 
A recent survey dealing with methodologies and tools to estimate the carbon footprint of ICT~\cite{Freitag2021} values between 2.1 et 3.9\% of the global GreenHouse Gas (GHG) emissions. 
In this variety of electronic and computing systems, data centers account for around two percent of the world's electricity consumption~\cite{masanet}, and surpass 10\% of the global electric consumption in some countries (e.g. Ireland)~\cite{iae_datacenters_impact}. 

Many researchers and decision makers believe that HPC will provide solutions to the warming/environmental crisis or could contribute to mitigate the crisis\footnote{https://sc24.supercomputing.org/2024/07/hpc-creates-sustainability/}. 
Large-scale climate simulations are the only way to understand global dynamics and gain a clearer view of future decisions. 
Optimizing components in the fields of transportation, energy, climate sciences, or intelligent buildings requires considerable computing power~\cite{rolnick}. 
HPC is also part of the problem, as it consumes a lot of electricity and natural resources (water, metals, etc.). 
In this paper, we develop an analysis to understand the carbon footprint of HPC. 
Our objective is to clarify the impacts of HPC from the perspective of energy consumption and carbon footprint. 
We briefly recall the empirical laws governing the evolution of HPC from performance and energy perspectives. The main contributions are the following.
\begin{itemize}
    \item We  study the performance and energy efficiency of HPC systems.
    \item We inestigate their effects and impacts for the lifespan of HPC systems.
    \item We derive an estimation of carbon emissions of HPC within the horizon 2030.
\end{itemize}

\section{Background}
\label{sec:background}

The Top500 ranking\footnote{https://top500.org/} is popular in the HPC community. Established in 1993, its objective is to evaluate and rank the world's most powerful HPC systems twice a year. It initially focused only on computational performance but evolved to include other system characteristics such as architecture, memory, power efficiency, and processor specifications. Top500 ranks systems biannually based on the benchmark of \textit{Linpack}~\cite{Linpack}. Linpack is a software library for linear algebra developed in the 1970s by the authors of Top500. A program is supplied by Top500 to be run on the HPC system. It requires a number of floating point operations during its execution.
This number divided by the runtime gives the metric $R_{max}$ in FLOPS. More than fifty lists have been published, including more than 10,000 HPC systems from more than 2,800 institutions around the world. 

Recognizing the increasing importance of energy efficiency, the Green500~\cite{Scogland2011} list was introduced in 2009. It provides a rearranged ranking to the Top500 list, based on energy efficiency (number of FLOPS/W). Officially added to the Top500 website in 2013, the Green500 list follows the same update cycle.

The main advantage of the Top500 is that it provides a stable view of the same features collected over many years and is thus a valuable source for measuring the evolution of HPC architectures. However, Top500 systems included in the list are submitted on a voluntary (not contractual) basis. The performance are self-valuated and self-reported with almost no independent checks. Despite its broad coverage, many systems are not included, for geopolitical or economical reasons, or simply because their companies have no interest in publishing them. Additionally, the Top500 list is oriented towards large scale HPC systems. 
\section{Related Works}
\label{sec:relatedworks}



Each Top500 list is released every year in May during the ISC High Performance conference and in November at SuperComputing.
It is an occasion to observe the evolution of trends in HPC. 
Regarding technology, Top500 authors published a study in 2015~\cite{Strohmaier2015} using the evolution of $R_{max}$ 
over time to evaluate the increase of processor capacity compared to Moore's law\footnote{A widely adopted observation stating that the number of transistors in an integrated circuit doubles about every two years.}~\cite{Khan2021,Subramaniam2013}. It describes key break points in the evolution that highlight a slowdown starting from 2008. 
This study is restricted to homogeneous systems and was published before the era of GPUs. 
Milojicic et al.~\cite{Milojicic2021} studied the evolution trend of HPC from an architectural point of view. They showed that HPC evolved to more customization. They also highlight the performance slowdown mentioned in~\cite{Strohmaier2015} but attribute it to the end of Dennard scaling\footnote{Dennard scaling envisions a physical limit of power per die area that restrains the performance gain due to increasing density of transistors}.
None of the above works really tackles the energy efficiency and its impact on the overall evolution of HPC systems. 
Khan et al.~\cite{Khan2021} conducted an analysis on the architectural trends of the Top500. They compare homogeneous and heterogeneous systems in terms of performance and power consumption. It was mostly focused on the period 2009-2019 that includes the diffusion of GPU in HPC. In~\cite{projection2011}, authors made performance and energy consumption projections up to 2024, claiming that only heterogeneous systems will be able to reach exascale but with an energy consumption of up to 100 MW.

Green500 was introduced to
study the evolution of the energy efficiency of HPC systems, that is, the ratio of $R_{max}$ over the power consumption. 
Scogland et al.~\cite{Scogland2011,Subramaniam2013} described the evolution of the Green500 metric over the first three years to identify the design aspects that contribute to greater power consumption in the objective of exascale. 
It introduces a new (holistic) metric to give a unified information about the energy efficiency/performance of Green500 called the EXASCALAR. 
This question of correlation between performance and energy efficiency is crucial for various studies. 
Khan et al.~\cite{Khan2021} evaluated the evolution of the Pearson’s correlation
coefficient of both lists during the period 2009-2019. 

Mair et al.~\cite{Mair2015} analyzed the energy efficiency of homogeneous and heterogeneous systems based on Green500, and proposed a new energy efficiency metric to avoid bias induced by system sizes.
Hsu et al. mentioned the same metric in~\cite{Hsu2016}. 
They provide a 10-year retrospective on the evolution of energy efficiency metrics to evaluate HPC systems (mainly the Power Usage Efficiency and Performance-Power Ratio). 
It also highlights the main issues that need to be addressed in terms of evaluation metrics and measurements.
Fraternali et al.~\cite{Fraternali2018} took a closer look at the impact of heterogeneity on performance and variability of energy consumption by conducting their experiments directly on one of the top system in Green500. Although their study does not address trend analysis, they were able to propose some guidelines (hardware-wise) to build sustainable HPC systems.
Gao et al.~\cite{Gao2016} relayed on Top500 and Green500 to study the influence of design and architectural parameters on the Linpack scores in terms of performance and energy efficiency. Then, they identified some development trends in HPC system design.






\section{The Top500 revisited}
\label{sec:experiments}

This section reports the multiple experiments and analysis performed on the Top500 and Green500 datasets after being preprocessed (i.e filtered by removing any incomplete items or outliers and unified in terms of measurement units). Experiments were produced with a \textit{Python3.8} kernel and take around 30 seconds to reproduce. The code used to produce the different figures, along with all the steps to replicate these experiments, is available on Github\footnote{\url{https://github.com/aBenhari/Green500-analysis.git}}.
We investigate, through these experiments, the performance and energy trends of HPC systems.


\subsection{Performance Evolution}
\label{subsec:perf}

\begin{figure}[t]
\centerline{
\includegraphics[width=1\textwidth]{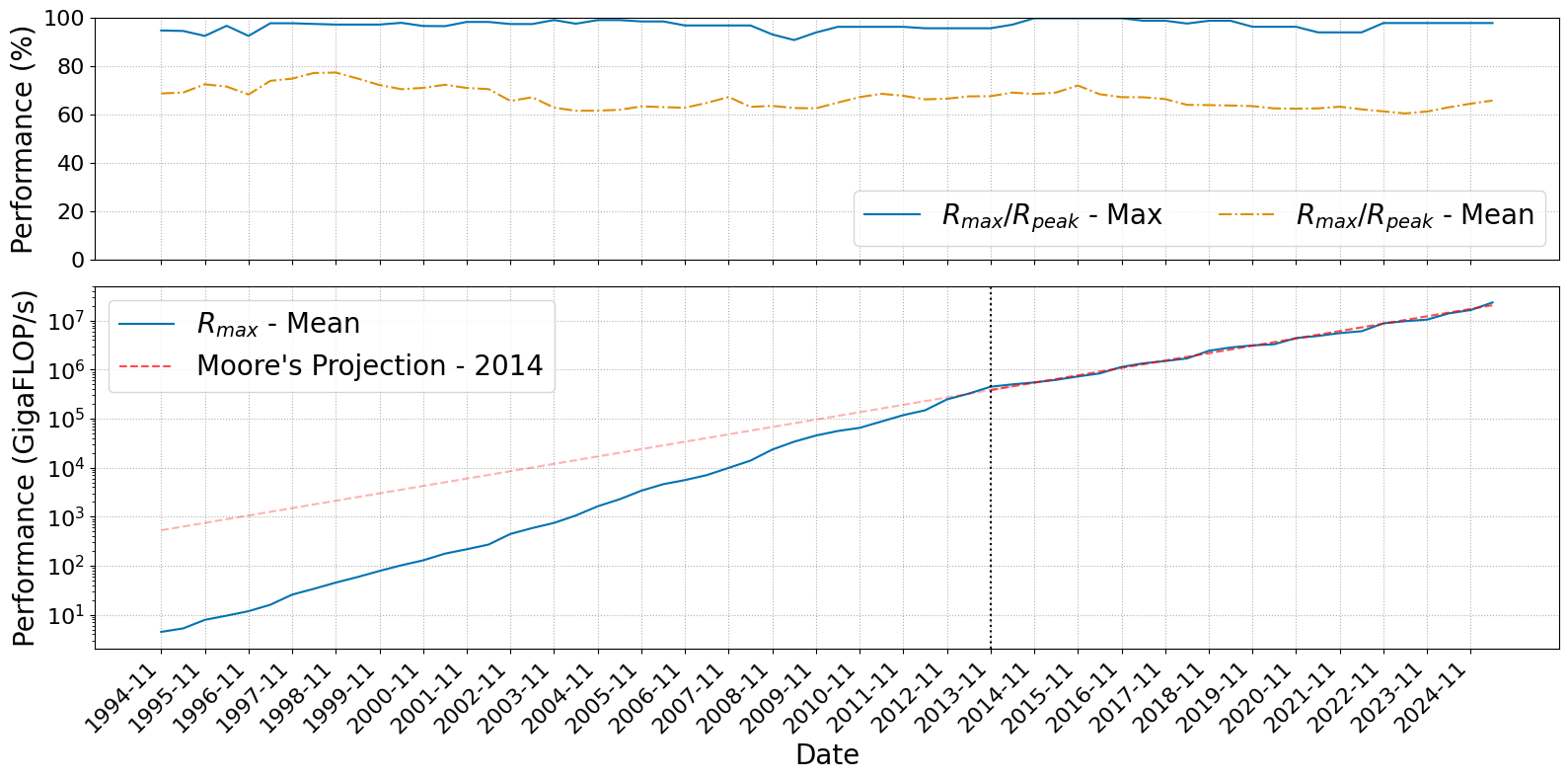}}
\caption{The evolution of the HPC system's performance metric ($R_{max}$) compared to Moore’s law projection from 2014 (\textit{bottom}). And the average performance ratio between the Linpack $R_{max}$ and the theoretical $R_{peak}$ for the {Top500} systems for the best system for this criteria (Max) and in average on the list (Mean)(\textit{top}).}
\label{fig:perf_evo_and_ratio}
\end{figure} 

The performance growth of HPC systems has been widely studied in the literature, especially in terms of $R_{max}$ evolution~\cite{Strohmaier2015, projection2011}. Figure~\ref{fig:perf_evo_and_ratio} (\textit{Bottom}) shows the evolution of $R_{max}$ (on y-axis) over time (every 6 months) illustrating this trend. We notice an increase following Moore's law with a clear break point in 2013 where the curve is starting to slow down but remains exponential (it fits exactly Moore's law). Although this pattern is well documented, our work focuses on performance evolution to analyze its dynamics and to compare it with the evolution of energy efficiency and carbon footprint of these systems.

Since the $R_{max}$ performance evolution on Top500 is purely driven by technology, similar trends are expected on other benchmarks than Linpack, such as HPCG (from Top500). However, the improvement ratio varies significantly across these benchmarks. To assess the impact of this progress, we analyze the ratio between the maximum performance reached using the Linpack benchmark ($R_{max}$) and the maximum theoretical performance $R_{peak}$, based on the specification of the computing components.

In Figure~\ref{fig:perf_evo_and_ratio} (\textit{Top}), the average performance percentage in the Top500 has decreased slightly, reflecting the increasing difficulty of utilizing these systems to their full potential, due to the growing heterogeneity of the hardware and the complexity of the infrastructure. However, the top-ranked systems remain efficient in this benchmark. This efficiency does not necessarily reflect their real-life performance, as they are highly optimized for Linpack benchmark (and similar workloads), thus limiting their versatility for other HPC applications.



\subsection{Turnover of HPC Systems}
\label{subsec:lifespan}



\begin{figure}[t]
\centerline{\includegraphics[width=0.9\textwidth]{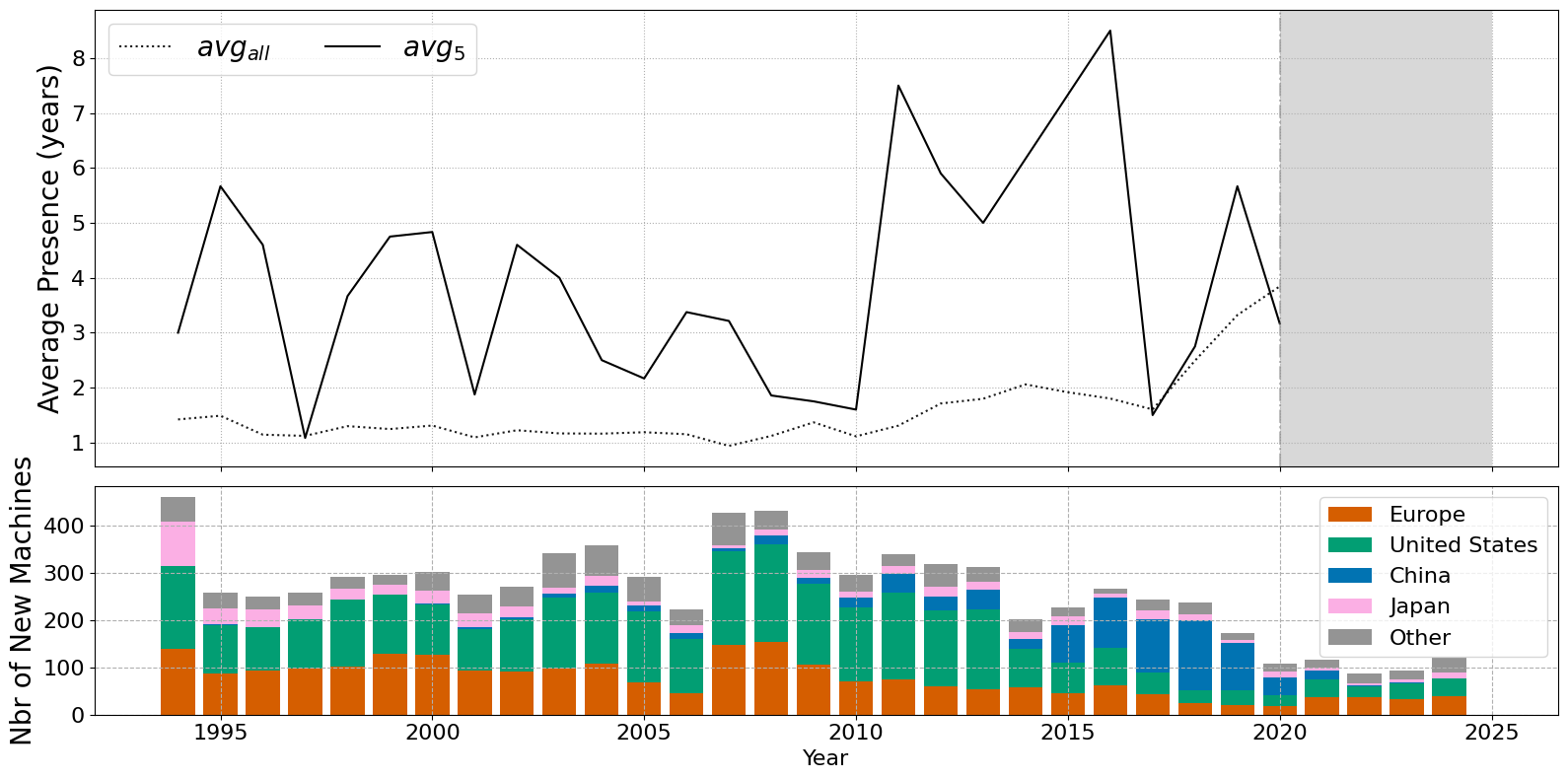}}
\caption{Average presence time (in years) per first apparition in \textit{Top500} for the first $5$, $50$, and $500$ systems. (The grey area represents the time frame of actual systems.) [\textit{Top}]. Number of newly registered HPC systems in the \textit{Top500} by region [\textit{Bottom}].}
\label{fig:avg_presence_and_arrival}
\end{figure}

In order to evaluate the turnover dynamics of HPC systems, we analyze their presence time in the Top500 list as a reference. We calculated that the listed systems remain on the list for $1.4$ years on average, with only the top ranked systems achieving longer presence periods. Figure~\ref{fig:avg_presence_and_arrival} (\textit{Top}) depicts the average presence time of HPC systems in the Top500 by rank group. It highlights a significant decrease in system turnover after 2010, especially for high-rank systems. This trend is correlated with fewer new systems entering the Top500 each year due to various external influences, such as geopolitic or fundings.

Although the Top500 provides useful insights about system longevity, it does not depict the full life cycle of HPC systems. The typical lifespan of HPC systems typically ranges from 5 to 7 years, where many of them are upgraded (e.g. \textit{Curie}\footnote{An HPC system developed by BULL and hosted at \href{https://www-hpc.cea.fr/fr/Joliot-Curie.html}{CEA}. It was ranked 9th in Top500 list in November 2011}), repurposed, or decommissioned. Figure~\ref{fig:avg_presence_and_arrival} (\textit{Bottom}) illustrates a decline in newly listed systems per year, stabilizing at 100 since 2020. This decline reflects a potential reduced pace of hardware replacement. Although the lifetime of an HPC system remains pertinent, it is important to assess that its environmental impact extends beyond operational use~\cite{labos1.5}.

A broader environmental impact assessment requires analyzing the operational (i.e. energy consumption) cost of HPC systems, but also the manufacturing and upgrade costs in terms of carbon footprint. In later sections, we detail a full carbon footprint LCA (Life Cycle Assessment) for selected HPC systems, including manufacturing costs, hardware upgrades, and usage impact. This approach provide an accurate evaluation of the total environmental costs of HPC.

\subsection{LCA of Specific Systems}

In this section, we investigate the manufacturing impact of selected HPC systems that were listed in the Top500 list during a maximum lifetime of 7 years.

The carbon footprint estimation for the \textit{Curie} and \textit{Fugaku}\footnote{A system developed by \href{https://www.r-ccs.riken.jp/en/fugaku/}{RIKEN} in the Top500 from June 2020 to June 2022.} systems were calculated using emissions data for similar hardware components, giving a broad estimate of the global carbon emissions of these systems. The manufacturing emissions were estimated using Baolin Li et al.'s work~\cite{tot_carb_hpc}, which indicates that HPC server CPUs have an embodied footprint of 250 kg $CO_2eq$ per node. Applying this to Curie’s \textit{Skylake} CPUs (1656 nodes, 28 CPUs per node), \textit{Irene Rome} CPUs (2292 nodes, 28 CPUs per node), and \textit{Fugaku’s ARM} processors (158 976 nodes, 48 CPUs per node) results in 602.8, 834.3, and 99,201 $tCO_2eq$, respectively. For \textit{NVIDIA V100} GPUs (128 cores), we used estimates provided by the \textit{NVIDIA Volta Whitepaper}~\cite{nvidiaVolta}, which adds 32 $tCO_2eq$. For memory and interconnect emissions we estimate 200 $kgCO_2eq$ per 64 GB of DDR4 and 5 $kgCO_2eq$ per GB of HBM2, resulting in 3790 $tCO_2eq$ for Curie DDR4 and 25436 $tCO_2eq$ for \textit{Fugaku} HBM2 in total. For Infiniband networking emissions, we approximated them at 50 $kgCO_2eq$ per node, contributing 7949 $tCO_2eq$ for \textit{Fugaku} and 82.8 $tCO_2eq$ for \textit{Curie}~\cite{tot_carb_hpc}. 

\subsubsection{Manufacturing Impact -}

Manufacturing contributes significantly to the total carbon footprint of HPC systems, sometimes reaching the carbon footprint generated by their use throughout their lifetime. As shown in Table~\ref{tab:carbon_emissions_fugaku}, the production of \textit{Fugaku} generated approximately 142,761 tons of $CO_2eq$ ($tCO_2eq$). The fabrication of CPU components accounts for 99,201 $tCO_2eq$, memory production adds 35,610 $tCO_2eq$, while interconnect and other hardware components contribute to roughly 7,950 $tCO_2eq$.
Similarly, the initial deployment of \textit{Curie} (Table \ref{tab:carbon_emissions_curie}) resulted in 3,342 $tCO_2eq$, mainly from CPU and memory components. Despite the difference in scale, both HPC systems share similar patterns of carbon emissions distribution, where CPUs represent the largest source of emissions, followed by memory and network components.

\begin{table}[t]
\centering
\caption{Estimated Carbon Emissions for \textit{Fugaku} (in $tCO_2eq$).}
\label{tab:carbon_emissions_fugaku}
\resizebox{\textwidth}{!}{\begin{tabular}{@{}llcc@{}}
\toprule
\textbf{Section}             & \textbf{Component}         & \textbf{Quantity}         & \textbf{Emissions (tons $CO_2eq$)} \\ \midrule
\multirow{3}{*}{\textbf{Manufacturing}}  
                             & CPUs                       & 158,976 nodes             & 99,201 \\ 
                             & Memory (HBM2)              & 5,087,232 GiB             & 35,610 \\ 
                             & Interconnect + Other       & 158,976 nodes             & 7,949 \\ \cmidrule{2-4}
\textbf{Total (Manufacturing)} &                           &                           & \textbf{142,761} \\ \midrule
\textbf{Usage (7 years)}     & Energy Consumption         & 1,713,600 MWh             & 376,992 \\ \midrule
\textbf{Total}         &                            &                           & \textbf{519,753} \\ \bottomrule
\end{tabular}}
\end{table}

\begin{table}[t]
\centering
\caption{Estimated Carbon Emissions for \textit{Curie} (in $tCO_2eq$).}
\label{tab:carbon_emissions_curie}
\resizebox{\textwidth}{!}{\begin{tabular}{@{}llcc@{}}
\toprule
\textbf{Section}             & \textbf{Component}         & \textbf{Quantity}         & \textbf{Emissions (tons $CO_2eq$)} \\ \midrule
\multirow{3}{*}{\textbf{Manufacturing (Skylake)}}  
                             & CPUs                       & 1,656 nodes (28 CPUs/node)                & 602.8 \\ 
                             & Memory + Network           & 1,656 nodes               & 2,231 \\ 
                             & Other Components           & -                         & 508.0 \\ \cmidrule{2-4}
\textbf{Total} & & & \textbf{3,342.0} \\ \midrule
\multirow{3}{*}{\textbf{Manufacturing (Irene Rome \& V100)}}  
                             & CPUs                       & 2,292 nodes (28 CPUs/node)                & 834.3 \\ 
                             & GPUs                       & 128 GPUs                  & 73.6 \\ 
                             & Other Components           & -                         & 2,231 \\ \cmidrule{2-4}
\textbf{Total} & & & \textbf{3,139.1} \\ \midrule
\textbf{Usage (Skylake, 2 years)} & -  & -                         & \textbf{508.0} \\ \midrule
\textbf{Usage (Irene Rome \& V100, 5 years)} & -  & -                         & \textbf{1,863.0} \\ \midrule
\textbf{Total}         &                            &                           & \textbf{8,852.1} \\ \bottomrule
\end{tabular}}
\end{table}

The substantial difference between \textit{Fugaku} and \textit{Curie} carbon footprint comes from different factors other than the scale. \textit{Fugaku} (2020) was first listed in the Top500 with a performance of 442 PFLOP/s, requiring far more hardware (i.e. CPUs, memory, and interconnect). In contrast, \textit{Curie} (2011) has reached 1.86 PFLOP/s, 100 times lower than \textit{Fugaku}. The relative impact on electricity and manufacturing footprint of these systems varies. However, with a same electricity mix around 300 $gCO_2eq$, we observe in both systems that the carbon impact of manufacturing corresponds to the footprint of
one year of electricity consumption.
These observations suggest that manufacturing is a major contributor to the total impact of HPC systems, where operational efficiency has improved significantly. The impact of energy consumption can be reduced through renewable sources or power management; however, the carbon emission of hardware manufacturing is unavoidable. Extending the lifespan of systems and discarding unnecessary upgrades is crucial to reduce the impact of HPC.

\subsubsection{Hardware Upgrades Impact -}

In addition to initial manufacturing, many HPC systems undergo considerable hardware modifications during their lifetime. These upgrades result in additional carbon emissions. In the case of \textit{Curie}, the upgrade phase costs around 3,139.1 $tCO_2eq$, which nearly doubles the total carbon footprint of the system. While hardware upgrades extend the usability of existing systems, they also considerably increase their emissions.


\section{Beyond Top500}

\subsection{HPC usage}


In this section, we investigate the utilization patterns of some HPC systems, to develop a broader understanding about the importance of upgrades (either adding new partitions or setting up new clusters), emphasizing the inefficiencies related to underutilization. White upgrading systems are often conducted to meet growing demand, this process overlooks significant disparities between actual and expected usage patterns across different systems.

\begin{figure}[t]
\centerline{\includegraphics[width=0.9\textwidth]{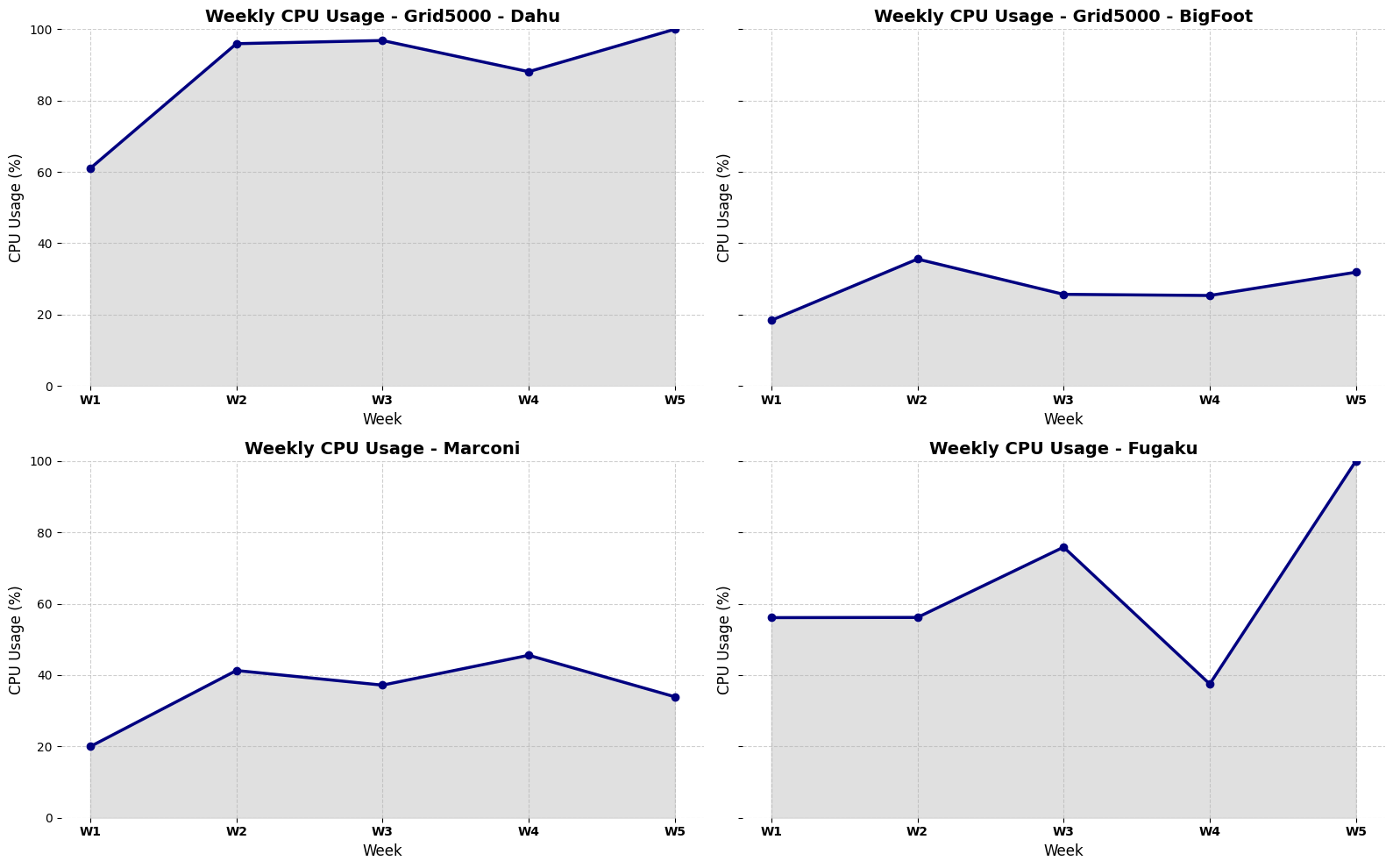}}
\caption{The percentage usage of three different HPC clusters (Fugaku, Marconi, Grid5000 - Dahu and Grid5000 - Bigfoot.) per time.}
\label{fig:machines_utilization}
\end{figure} 

Beyond our analysis of the Top500 lists, which primarily assess efficiency, we shift our focus to real-world utilization trends. To provide context for this analysis, we study four systems with different configurations and purposes: \textit{Fugaku}, \textit{Marconi100}\footnote{A system hosted by \href{https://www.hpc.cineca.it/systems/hardware/marconi100/}{CINECA} appeared in the Top500 in 2020 at the 11th position.}, \textit{Grid5000\footnote{\href{https://www.grid5000.fr/w/Grid5000:Home}{Grid5000} is a large-scale testbed composed of multiple HPC clusters in France, designed for research in parallel, distributed, and cloud computing.} - Dahu}, and \textit{Grid5000 - Bigfoot}. Both \textit{Fugaku} and \textit{Marconi} appeared in the Top500, thus representing systems tailored to perform large scale calculations. In contrast, \textit{Grid5000 - Dahu} and \textit{Grid5000 - Bigfoot} represent flexible systems used for tests and experimentation.

Figure \ref{fig:machines_utilization} illustrates the utilization rates of these systems over time. It highlights the periods of underutilization for each of them. For systems with persistent underutilization, upgrades aimed at improving performance or expanding capacity are unnecessary. This discrepancy underscores the need to carefully evaluate actual usage patterns before committing to upgrades, to avoid any unnecessary environmental impact.
This highlights the need of utilization driven evaluations (rather than relying only on performance metrics) when considering hardware upgrades. For example, a primary workload distribution assessment can help organizations make informed decisions, avoiding unnecessary replacement or expenditure.

\subsection{Energy Efficiency}
\label{subsec:energyefficiency}

Our purpose in this section is to highlight the progress made in terms of energy efficiency and to compare it with the performance. We focus on Koomey's law\footnote{An equivalent to Moore's law for the energy efficiency trend. He observed that the number of computations per joule of energy roughly doubles every 18 months.} as reference. 


\begin{figure}[t]
\centerline{\includegraphics[width=0.9\textwidth]{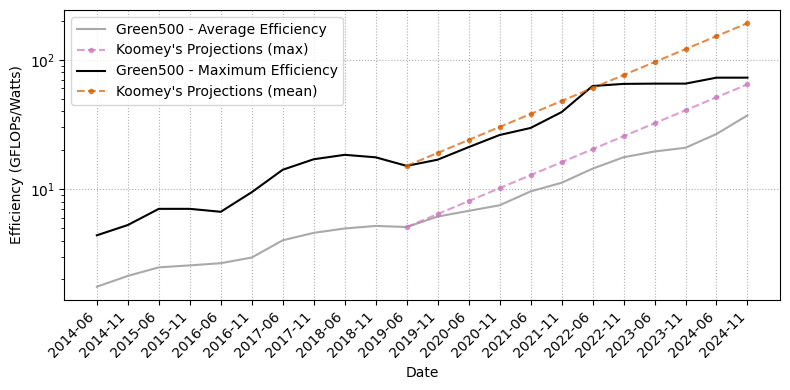}}
\caption{Maximum and Average \textit{Efficiency} of the \textit{Green500} 
HPC systems by date along with Koomey's law projections starting from 2019.}
\label{fig:Koomey}
\end{figure}

Figure~\ref{fig:Koomey} compares of energy efficiency evolution of HPC systems in GFLOPS/W (\textit{y-axis}) to Koomey's law projections (\textit{x-axis}).
We observe an increase in the energy efficiency, but at a lower rate than the performance. 
The maximum value increased from $4.5$ GFLOPS/W in the late 2013 to $65.39$ GFLOPS/W in 2023. However, this progress fails to follow Koomey's law with a doubling time higher than 1.5 years, especially at the beginning of the {Green500} history. It seemed to catch up during a few years after 2019 due to a rising interest in energy efficiency in the community and the emergence of GPU that are more energy efficient
as they have achieved higher positions in the Green500 ranking compared to homogeneous ones. However, the last lists have not showed significant improvement.

\section{Projection Scenarios}
\label{sec:projection}

We observed in Figure~\ref{fig:avg_presence_and_arrival} an increase in the lifetime of HPC systems in the Top500 over the last few years. However, this trend can be correlated to the decrease of the number of new systems per year. 
This number is fairly constant since 2020 around 100 new systems per year. This is consistent with a renewal of one fifth of the systems per year and then a mean lifetime around 5 years. Thus, we cannot expect the mean lifetime in Top500 to continue growing in the next years.

Considering performance, Figure~\ref{fig:perf_evo_and_ratio} indicates a clear trend since 2014 with an increase corresponding to Moore law. We can thus expect in 2030 a mean $R_{max}$ value around 130 PFLOPS.
This increase could be offset by an increase in efficiency following Koomey's law, but the top efficiency has stagnated in the last years, as observed in Figure~\ref{fig:Koomey}.



\begin{figure}[t]
\centerline{\includegraphics[width=0.9\textwidth]{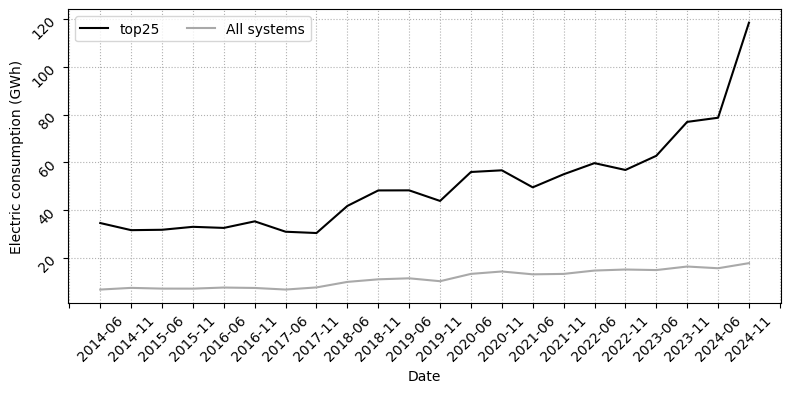}}
\caption{Average electric consumption (1 year) of the top 25 and all systems in the Top500 list over time.}
\label{top500_conso_co2}
\end{figure}

Figure~\ref{top500_conso_co2} shows the evolution of the electricity consumption of the first Top500 systems since 2014. It highlights the big difference between the top 25 systems and the following. Consumption is almost constant before 2018 and increases until 2023. 


\begin{figure}[h]
\centerline{\includegraphics[width=.8\textwidth]{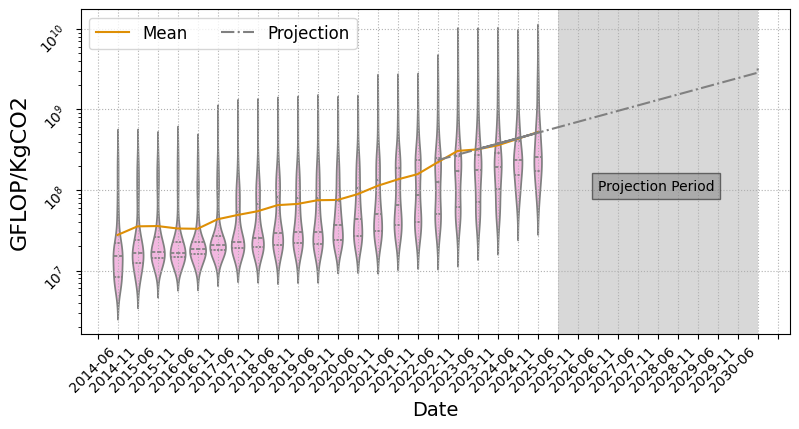}}
\caption{Evolution of the maximum performance in GigaFLOPS per kg\ch{CO2} in the Green500. 
The grey area represents the projection window.}
\label{gflops_per_kgco2}
\end{figure} 

Figure~\ref{gflops_per_kgco2} shows the evolution of carbon efficiency\footnote{The number of FLOP per kg of \ch{CO2}.} in Top50. For each date, the width of the shape shows the number of systems for a given efficiency. 
We observe a regular increase of the mean efficiency since 2014 with a doubling period of 2.83 years on average and a tendency to heterogeneity.
The first Top500 systems are close to the mean, while the head of Green500 increases faster.




Based on this analysis, we provide an estimate of the electricity consumption of HPC to compare with the roadmap for decarbonization.
Various scenarios are provided by IPCC~\cite{IPCC2022}.
The objective of the European Union Green Deal~\cite{europeancommissionDeliveringEuropeanGreen2021} is to reduce the GHG emissions by 55\% by 2030. 
By extrapolating from Figure~\ref{gflops_per_kgco2}, we envision the carbon efficiency to reach $1.64$ EFLOP per kg\ch{CO2} in 2030 compared to $0.35$ in 2022, which corresponds to 24.99\% per year of the value recorded in 2022. 
This is a rough estimate, but it shows great stability over a long period of time. 
In terms of carbon budget, how many FLOPS can be processed while respecting the European carbon reduction goal of 55\% with a planned power efficiency improvement of 537\%? If we take 2022 as a reference, then the total number of FLOPS without reduction is 5,37 more and 45\% of that number gives 2,41. To respect a 55\% carbon reduction goal, we can process more or less 2,5 times more FLOPS in 2030 than we did in 2022. A first step to improve this result is to lower the carbon intensity of the electricity that powers the Green500. This is not sufficient and HPC usage have to be reconsidered.

%
%


\section{Conclusion}
\label{sec:conclusion}

We studied the evolution of HPC using mainly data from the Top500.
In-depth analysis of both performance and energy efficiency has enabled us to better understand the environmental impacts of the domain
and highlights several important results.
(i) the performance and energy efficiency increase has diminished between 2014 and 2018, but keeps increasing and (ii) energy efficiency does not fit anymore Koomey's law and the total energy consumption of Top500 systems keeps increasing in the recent years. 
We have proposed a prospective study up to 2030. 
The consumption of HPC systems continues to rise when GHG emissions need to be drastically reduced. 
However, multiple applications carried out on these large-scale systems may also help to propose solutions to mitigate the crisis. 
Today, IPCC experts consider that it is still time to act. 
HPC needs to participate by improving its footprint. 






\bibliographystyle{splncs04}
\bibliography{biblio.bib}

\end{document}